\documentclass[conference]{IEEEtran}
\IEEEoverridecommandlockouts
\usepackage{adjustbox}
\usepackage{soul}
\usepackage{comment}
\usepackage{amsmath}
\usepackage{float}
\usepackage{caption}
\usepackage{tabularx}
\usepackage[table,xcdraw]{xcolor}
\usepackage{pifont}
\usepackage{changepage}
\usepackage{gensymb}
\usepackage{fontawesome}
\usepackage{multirow} 
\usepackage[edges]{forest}
\usepackage{hyperref}
\usepackage{makecell}
\usepackage{amssymb}
\usepackage{booktabs}
\usepackage{tcolorbox}
\usepackage{tabto}
\usepackage{subcaption}
\usepackage{tikz}

\usepackage{graphicx}

\usepackage{verbatim}

\usepackage{algorithmic}
\usepackage{diagbox}
\usepackage{tikz}
\usetikzlibrary{shapes.geometric}
\usetikzlibrary{positioning}
\usepackage{tabularx}
\usepackage{ragged2e}
\usepackage{lipsum} 
\usepackage{amsmath}
\usepackage{amssymb} 
\usepackage{array}
\begin{document}
\bstctlcite{IEEEexample:BSTcontrol}



\title{The Value of Disagreement: A Bhattacharyya-augmented Object Selection Function for Collective Perception}

\author{
    Tengfei Lyu\textsuperscript{1}, Dirk Pesch\textsuperscript{1}, Aisling O'Driscoll\textsuperscript{1} \\
    \textsuperscript{1}nasc Research, School of Computer Science \& IT, University College Cork, Ireland.\\
    Email: t.lyu@cs.ucc.ie, \{dirk.pesch, aisling.odriscoll\}@ucc.ie
    \vspace{-2mm}
}
\maketitle
\begin{abstract}
Collective perception is regarded as a key enabler for connected and automated driving, as it extends situational awareness beyond the coverage of local sensors. However, the current European Telecommunications Standards Institute (ETSI) Collective Perception Service (CPS) specification relies mainly on dynamics and time triggered object inclusion rules, which may unintentionally promote the transmission of noisy sensor tracks and does not explicitly distinguish between the importance of objects or regions to the wider collective perception process when channel resources are constrained. To address this limitation, ETSI has introduced the concept of Value of Information (VoI)-based prioritisation, but has not yet formally specified VoI functions to calculate these scores, despite being under discussion. Recent work has attempted to address this gap by defining VoI requirements and implementing them using the Kullback-Leibler Divergence (KLD) and Expected Covariance Divergence (ECD) formulae. While promising, both functions exhibit potential limitations in safety-relevant object disagreement cases. KLD can be distorted by anisotropic uncertainty, leading to over- or under-inflated VoI scores, while ECD does not explicitly recognise object state disagreement. As a result, ECD may downgrade the transmission importance of an object where the ego vehicle and its surrounding neighbours/infrastructure have similar certainty yet conflicting estimates of the object state. Although such disagreement may be uncommon, it can be significant in safety critical scenarios and it is important that such disagreements can be shared so that a suitable decision is taken. This paper introduces a new CPS-specific VoI formulation and evaluation, namely the Bhattacharyya-enhanced Covariance Divergence (BCD) VoI object-selection function. BCD integrates the established Bhattacharyya coefficient into the covariance-driven structure of existing VoI functions. This results in bounded disagreement-aware VoI scores that enable more balanced object inclusion, prioritising disagreement under payload constraints. The impact of BCD on receiver-side estimation and the allocation of transmission opportunities among competing objects is evaluated. Controlled simulations show that BCD achieves the lowest aggregate receiver error in payload-constrained scenarios. It improves the treatment of safety-relevant disagreement relative to ECD while avoiding the object starvation observed with KLD.
\end{abstract}

\begin{IEEEkeywords}
CPS; Value of Information (VoI); Sensor Data Sharing, V2X.
\end{IEEEkeywords}

\section{Introduction}
\label{sec:introduction}
The Collective Perception Service (CPS) disseminates object and perception region information through Collective Perception Messages (CPMs), allowing vehicles and roadside units (RSUs) to extend their perception of the environment~\cite{etsiintelligent} beyond the coverage of their own sensors. This is particularly important in occluded scenarios, where an Intelligent Transport System Station (ITS-S) may not be able to directly perceive a safety-relevant road user or object. Currently, CPM generation is primarily triggered by dynamics and time object inclusion rules. These rules rely on criteria such as whether an object is newly detected, has changed sufficiently in position or motion state, or has not been reported for a defined time interval~\cite{etsiintelligent}. Although this design is simple, interoperable, and easy to implement, dynamics-based inclusion can be sensitive to sensor noise and tracking instability~\cite{Lyu2025Filtering}. A noisy or weakly converged object track may repeatedly cross kinematic thresholds to be transmitted, even if its contribution to collective perception is limited. This consumes CPM payload and channel resources and may cause more relevant objects to be omitted when not all candidates can be included~\cite{thandavarayan2020cooperative,thandavarayan2020redundancy}.

The object-selection problem is therefore not only a question of reducing CPM size but also of deciding which locally perceived objects are sufficiently reliable and useful to improve collective scene understanding. A promising way to address this is to prioritise objects according to their \textit{Value of Information (VoI)}. ETSI already specifies the concept of VoI for prioritising high-value objects in multi-channel operation and update rate control mechanisms~\cite{etsiintelligent}. However, concrete VoI-based object inclusion requirements and recommended implementations are not yet formally specified. Moreover, existing prioritisation or update rate control mechanisms are typically applied only after an object has already satisfied the dynamics-based inclusion rules. The candidate set of objects entering the prioritisation stage can therefore remain shaped by sensor noise, tracking instability, or temporary measurement fluctuations.

Prior work has specified VoI-based object inclusion requirements based on local sensing accuracy and the relevance of sharing a perceived object with the network~\cite{lyu2025accuracy}. These form a subset of the requirements currently under discussion at ETSI. Thus, a VoI implementation function should favour objects that are accurately estimated by the transmitting ITS-S, and whose transmission is expected to improve the receiver-side view, either because the object is uncertain to others or because the local and exchanged estimates disagree. Two promising VoI functions have been proposed based on \textit{Kullback-Leibler Divergence (KLD)} and \textit{Expected Covariance Divergence (ECD)}~\cite{lyu2025accuracy}. These functions were further evaluated by Wolff et al.~\cite{wolff2025Uncertainty}. Conceptually, both functions compare what the transmitting ITS-S  knows with what it believes the surrounding network is aware of through V2X information. 
KLD also attempts to prioritise information where stations disagree on the estimated object state, so that receivers are aware of potentially safety-relevant inconsistencies. Such disagreement may be uncommon under normal operating conditions, but it can be important in edge cases. For example, two ITS-S' may estimate a vulnerable road user at slightly different lateral positions: one estimate may place the VRU inside a drivable lane and potentially in conflict with the receiver’s trajectory, while another may place the VRU on a footpath. In this case, sharing the disagreement is useful so the receiver does not treat the object state as fully resolved. Existing VoI functions do not fully address disagreement. ECD does not consider it at all, and KLD attempts to address it via a Mahalanobis-distance term, but this is scaled by the inverse covariance of the estimated network knowledge. As a result, KLD may over- or under- estimate safety-relevant conflicts. Importantly, while it is desirable to share object state disagreement, the VoI functions should not promote the over-sharing of inaccurate object states, e.g. if the local track is affected by association error, calibration bias, delayed sensing, or another perception fault, repeatedly transmitting this information may propagate erroneous information. The design goal is therefore to recognise and share disagreement without allowing it to dominate the CPM object VoI score ranking. A suitable VoI function should raise the priority of safety-relevant disagreement while preserving transmission opportunities for other useful objects.

This paper proposes a \textit{Bhattacharyya-enhanced Covariance Divergence (BCD)} VoI function. It retains the covariance-driven structure of existing VoI functions, but replaces the Mahalanobis-distance disagreement term in KLD with a bounded overlap-based term derived from the Bhattacharyya coefficient~\cite{bhattacharyya1943aspect}. It prioritises the sharing of disagreements while avoiding the overly aggressive transmission behaviour that can occur with KLD. In this way, BCD provides more balanced behaviour: overcoming the disagreement unaware characteristics of ECD, but with more balanced and less frequent transmissions than KLD. The paper contributions can be summarised as:

\begin{itemize}
    \item Description of the conditions/state under which a safety-relevant CPS object selection disagreement can occur.
    \item An analysis of why existing VoI functions may under- or over-prioritise such states.
    \item  A bounded disagreement-aware VoI function \textit{BCD} is proposed that better handles these conditions based on the \textit{Bhattacharyya coefficient}. Simulation results show that when the channel is congested BCD provides more balanced treatment of safety-relevant disagreement.
\end{itemize}

\section{Background and Motivation}
\label{sec:background}
This section introduces the operation of the proposed VoI functions in meeting the accuracy, network relevance and object state disagreement (optional) requirements under discussion as part of the ETSI CPS task force. This includes summarising important VoI function terminology and the states used to derive VoI scores (Section~\ref{subsec:lem_vei_states}), the conditions under which object state disagreement occurs (Section~\ref{subsec:state_b_causes}) and how the existing proposed VoI functions fail to adequately address object state disagreement (Section~\ref{subsec:existing_voi_state_b}).  

\subsection{Operation of existing VoI functions}
\label{subsec:lem_vei_states}
To derive VoI scores for sensed objects based on their accuracy, relevance to the network and object statement disagreement, ECD and KLD represent the knowledge of a transmitting ITS-S through two tracking estimation models~\cite{lyu2025accuracy,wolff2025Uncertainty}: the \textit{Local Environment Model (LEM)} and the \textit{V2X Exchanged Information (VEI)}. The LEM describes the object state estimated only from the transmitter's own local sensors. The VEI describes the object state that the transmitter believes is already known to other ITS-S' via previously transmitted and received V2X messages. For object $o$ at time step $k$, the LEM estimate is represented as:
\begin{equation}
    \left( \boldsymbol{\mu}^{\mathrm{LEM}}_{o,k},
    \boldsymbol{\Sigma}^{\mathrm{LEM}}_{o,k} \right),
    \label{eq:lem_estimate}
\end{equation}
where $\boldsymbol{\mu}^{\mathrm{LEM}}_{o,k}$ is the local mean object state and $\boldsymbol{\Sigma}^{\mathrm{LEM}}_{o,k}$ is the associated covariance matrix. Similarly, VEI is
\begin{equation}
    \left( \boldsymbol{\mu}^{\mathrm{VEI}}_{o,k},
    \boldsymbol{\Sigma}^{\mathrm{VEI}}_{o,k} \right),
    \label{eq:vei_estimate}
\end{equation}
where $\boldsymbol{\mu}^{\mathrm{VEI}}_{o,k}$ and $\boldsymbol{\Sigma}^{\mathrm{VEI}}_{o,k}$ denote the mean object state and covariance of the V2X-derived estimate. An object's value for transmission increases when the local LEM estimate increases in reliability and the corresponding VEI estimate is uncertain, outdated, or inconsistent. This interpretation connects the VoI of an object to both local sensing accuracy and the expected benefit that it might bring to neighbouring ITS-S'.
\begin{figure}[t]
\centering
\subfloat[\textbf{State A:} The LEM is more certain than the VEI.]{
  \includegraphics[width=0.45\linewidth]{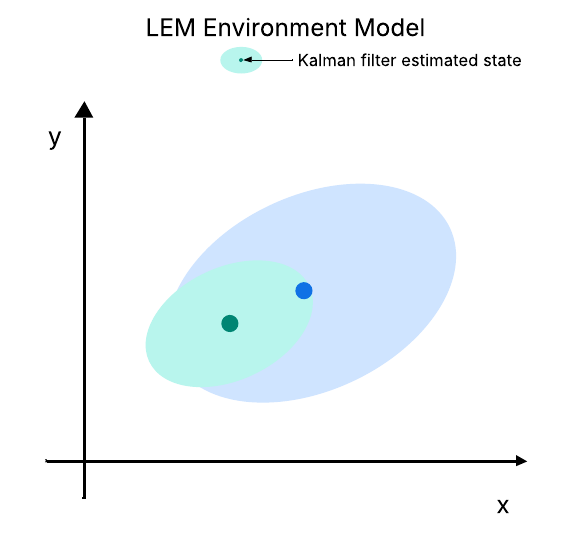}%
  \label{fig:state_a}%
}
\hfill
\subfloat[\textbf{State B:} Similar uncertainty, but strong object state conflict.]{
  \includegraphics[width=0.45\linewidth]{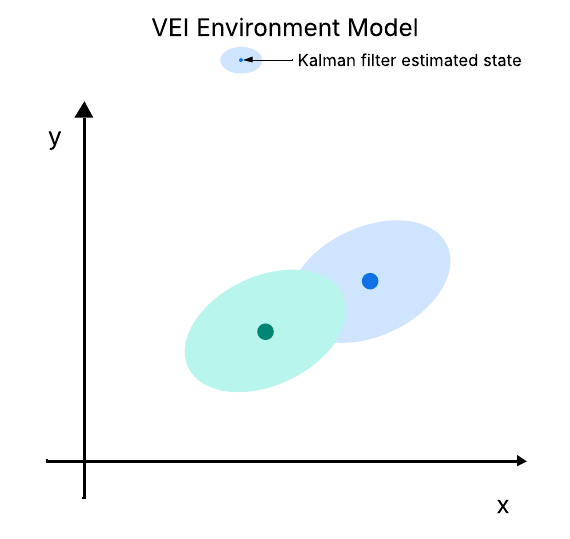}%
  \label{fig:state_b}%
}
\\[0.5ex]
\subfloat[\textbf{State C:} Similar uncertainty, and strong state agreement.]{
  \includegraphics[width=0.45\linewidth]{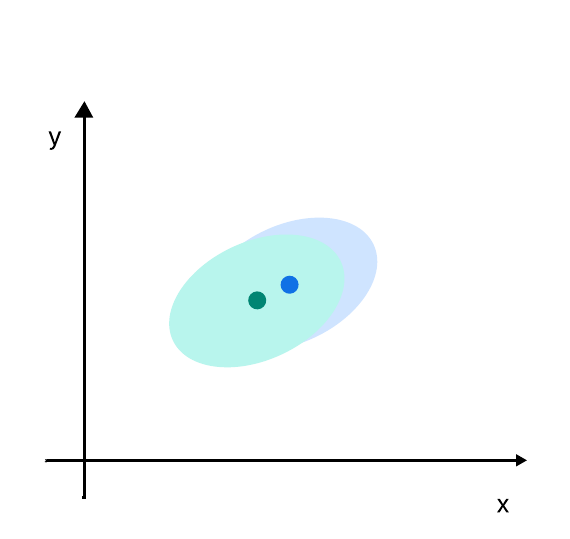}%
  \label{fig:state_c}%
}
\hfill
\subfloat[\textbf{State D:} The VEI is more certain than the LEM.]{
  \includegraphics[width=0.45\linewidth]{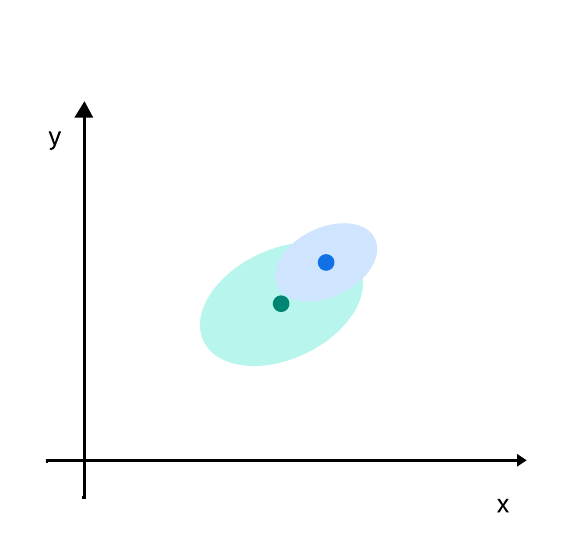}%
  \label{fig:state_d}%
}
\caption{Four states that contrast LEM and VEI tracking estimates to derive a VoI score for the same object. The states differ in relative uncertainty and estimated mean object states.}
\label{fig:Four_state_interpretation}
\vspace{-2mm}
\end{figure}

The relationship between the LEM and VEI can be described using four representative \textit{states}, as illustrated in Fig.~\ref{fig:Four_state_interpretation}. State A occurs when the LEM is more certain than the VEI. In this case, the object is accurately perceived locally but remains uncertain in the V2X-exchanged information, if transmitted, it is expected to improve the network-side knowledge. State D is the opposite case, where the VEI is more certain than the LEM. In this case, the object is already well represented in the exchanged information, so its VoI should be low.

The key distinction in this paper is between State B and State C. In both states, the LEM and VEI have comparable uncertainty. The distinction is whether their mean object states agree. In State C, the two estimates overlap and describe the object consistently. In State B, the two estimates have similar uncertainty but different mean states. This means that the locally perceived object state and the V2X-derived object state are inconsistent. Under this controlled interpretation, the desired VoI ordering is:
\begin{equation}
    \mathrm{State~A} > \mathrm{State~B} > \mathrm{State~C} > \mathrm{State~D}
    \label{eq:state_ordering}
\end{equation}
This ordering should not be interpreted as a universal ranking across all traffic scenes, since real objects may differ in object type, distance, absolute uncertainty, occlusion risk, and safety relevance. VoI scores may therefore vary as LEM and VEI uncertainty estimates change. Rather, Eq.~\eqref{eq:state_ordering} expresses the intended behaviour under identical, controlled LEM conditions. 

\subsection{Conditions under which Object State Disagreement occurs}
\label{subsec:state_b_causes}
State B should not be interpreted as a normal steady-state condition. In a collective perception system, the LEM and VEI are maintained simultaneously on the same ITS-S and should converge towards a consistent representation of the same object. A persistent State B condition, therefore, indicates an inconsistency between local perception and V2X-derived perception. Importantly, State B does not reveal which estimate is correct. It only indicates that the LEM and VEI have comparable uncertainty but disagree in their mean object state. This makes State B safety-relevant, but also ambiguous: a VoI function should raise its priority without blindly assuming that the LEM is more reliable than the VEI.

\begin{figure}[t]
    \centering

    \begin{subfigure}[t]{0.48\columnwidth}
        \centering
        \includegraphics[width=\linewidth]{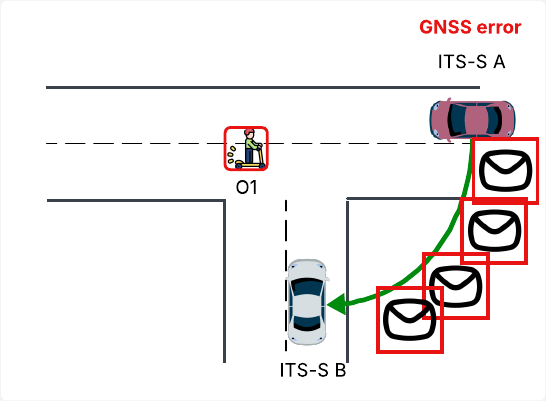}
        \caption{Transmitter positioning bias.}
        \label{fig:gnss_error}
    \end{subfigure}
    \hfill
    \begin{subfigure}[t]{0.48\columnwidth}
        \centering
        \includegraphics[width=\linewidth]{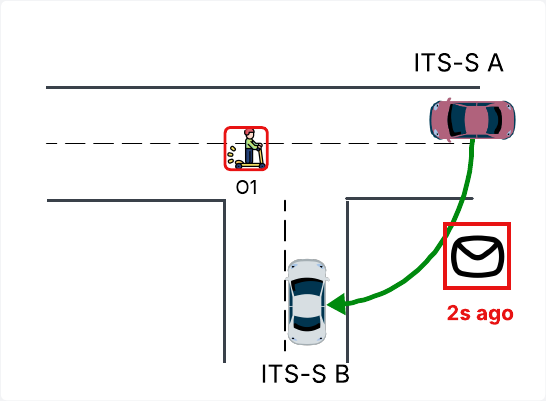}
        \caption{Outdated V2X update.}
        \label{fig:stale_message}
    \end{subfigure}

    \vspace{0.4em}

    \includegraphics[width=0.32\columnwidth]{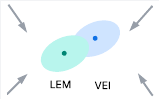}

    \vspace{-0.4em}

    \begin{subfigure}[t]{0.48\columnwidth}
        \centering
        \includegraphics[width=\linewidth]{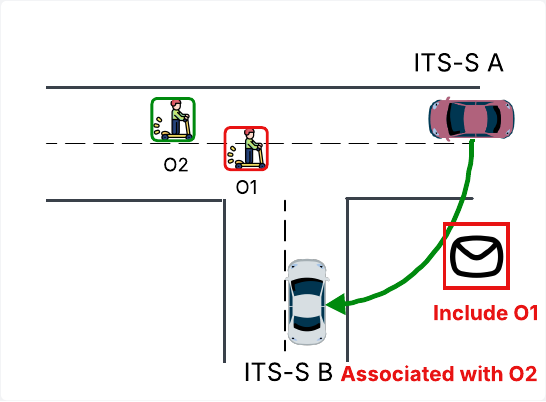}
        \caption{Association ambiguity.}
        \label{fig:association_error}
    \end{subfigure}
    \hfill
    \begin{subfigure}[t]{0.48\columnwidth}
        \centering
        \includegraphics[width=\linewidth]{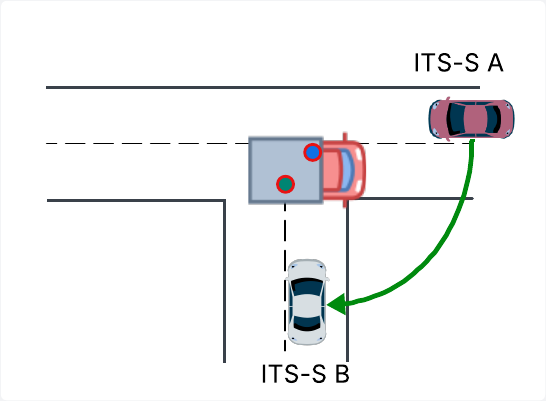}
        \caption{Sensor calibration or viewpoint bias.}
        \label{fig:viewpoint_bias}
    \end{subfigure}

    \caption{Representative causes of object state disagreement. In all four cases, the LEM and VEI may remain comparably certain but disagree in their estimated object state.}
    \label{fig:state_b_causes}
    \vspace{-2mm}
\end{figure}

Fig.~\ref{fig:state_b_causes} illustrates four representative causes of State B. These causes are not exhaustive, but they show how LEM and VEI disagreement can arise in practical collective perception systems. On analysing an enriched real-world inD dataset~\cite{inDdataset}, the rarity of State B can be observed. Under the assumption that all vehicles use identical sensor configurations and that the communication channel is perfect, State B objects account for only around $0.4\%$ to $0.5\%$ of all object states during the $900$~s simulation. This confirms that State B is not expected to dominate normal CPM generation. Instead, it represents a low-frequency but potentially safety-relevant edge case that a VoI function should handle correctly when it occurs.
A first cause is shown in Fig.~\ref{fig:gnss_error}. A transmitting ITS-S (ITS-S A) may report perceived objects using a biased reference position, i.e. Global Navigation Satellite System (GNSS) error. In this case, the objects reported through V2X shift the receiver's VEI object estimate further from the LEM, even though both estimates refer to the same physical object. If both estimates remain confident, this produces a State B condition. The second cause is a delayed V2X update, as shown in Fig.~\ref{fig:stale_message}. The VEI is updated only when new V2X information is received. If the most recent update is delayed, lost, or generated at a lower rate, the VEI may still represent an older object position while the LEM has already tracked the object at its current position. If the uncertainty of the stale VEI estimate is not increased sufficiently, the LEM and VEI can appear similarly certain but disagree in their mean object states. The third cause is association ambiguity, as shown in Fig.~\ref{fig:association_error}. In dense scenes, two nearby objects may be difficult to distinguish, especially around junctions, crossings, or occluded areas. A received V2X object may then be associated with the incorrect local track. This can produce a confident yet incorrect pairing of LEM and VEI. The resulting disagreement is not caused by the physical motion of a single object, but by an incorrect association between the local and exchanged object states. The fourth cause is shown in Fig.~\ref{fig:viewpoint_bias} where different ITS-S' may observe the same object from different viewpoints or with different sensor configurations. Calibration errors, different fields of view, partial occlusion, or differences in object-shape estimation can shift the estimated object centre. If both ITS-S' remain confident in their own estimates, the LEM and VEI may have comparable covariance but limited overlap. 

\subsection{Limitations of Existing VoI Functions for Disagreement}
\label{subsec:existing_voi_state_b}
Representative VoI functions, KLD and ECD~\cite{lyu2025accuracy,wolff2025Uncertainty}, both compare their LEM and VEI estimates of a perceived object to derive VoI scores. However, their behaviour differs significantly when handling State B states.

ECD, shown in Fig~\ref{fig:ecd_formula}, evaluates the covariance relationship between the LEM and VEI by examining the relative uncertainty structure and log-determinant ratio to determine accuracy and network relevance respectively. It does not directly evaluate the disagreement between the mean object states. This behaviour is useful in State A, where the LEM is more certain than the VEI, because transmitting the object can improve the network-side estimate. It is also relatively stable under noisy measurements, since small fluctuations in the estimated mean state do not directly increase the VoI score. However, this becomes a limitation in State B, where there is object state disagreement. When the LEM and VEI have comparable uncertainty, ECD cannot distinguish whether the two estimates agree or disagree. As a result, State B (disagreement) and State C (agreement) may receive similar scores.

KLD, as shown in Fig~\ref{fig:kld_formula}, addresses this limitation by including a mean disagreement term in the form of a Mahalanobis-distance component, shown in green, between the LEM and VEI mean object states. However, because this is weighted by the inverse VEI covariance, the same physical disagreement can receive different scores depending on the shape and orientation of the VEI uncertainty. Disagreement along a high-variance direction may be underweighted, whereas disagreement along a low-variance direction may be amplified and dominate the object ranking under  Decentralised Congestion Control (DCC) payload constraints. Thus, State B object state disagreement requires more balanced treatment in the derivation of VoI scores. This paper proposes a bounded disagreement-aware VoI formulation that can distinguish State B from State C without allowing disagreement to dominate the object VoI ranking.

\section{Proposed Bhattacharyya-enhanced Covariance Divergence (BCD) VoI function}
\label{sec:proposed_bakld}
The discussion in Section~\ref{sec:background} motivates that a VoI function for CPM object inclusion should not only reward local sensing accuracy and network-side uncertainty, but should also react to safety-relevant disagreement in a way that is \textit{proportionate} so as to not over or under amplify the VoI score of the disagreement nor starve other objects from being included in a payload constrained scenario. The proposed \textit{Bhattacharyya-enhanced Covariance Divergence (BCD)} function retains the covariance-driven structure of KLD, but replaces its Mahalanobis-distance term with a bounded overlap-based disagreement term.

\subsection{BCD VoI Function}
\label{subsec:bakld_definition}
\begin{figure}[t]
    \centering
    \captionsetup[subfigure]{font=footnotesize,skip=0pt}

    \begin{subfigure}[b]{0.84\linewidth}
        \centering
        \includegraphics[
            width=\linewidth,
            trim=0 4 0 4,
            clip
        ]{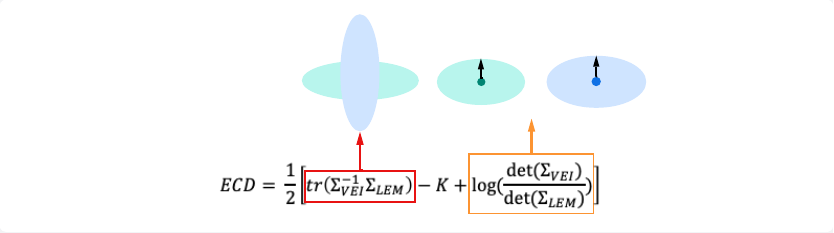}
        \caption{ECD}
        \label{fig:ecd_formula}
    \end{subfigure}
    \vspace{0mm}

    \begin{subfigure}[b]{0.84\linewidth}
        \centering
        \includegraphics[
            width=\linewidth,
            trim=0 4 0 4,
            clip
        ]{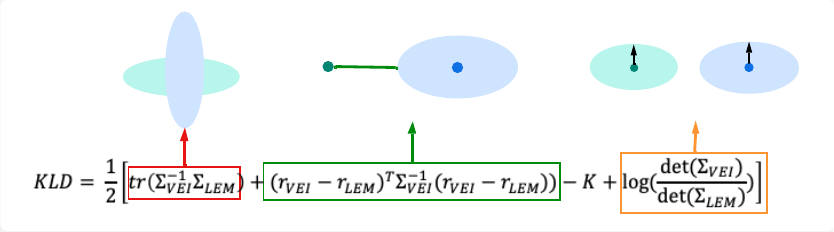}
        \caption{KLD}
        \label{fig:kld_formula}
    \end{subfigure}
    \vspace{0mm}

    \begin{subfigure}[b]{0.84\linewidth}
    \centering
    \includegraphics[
        width=\linewidth,
        trim=0 0 0 4,
        clip
    ]{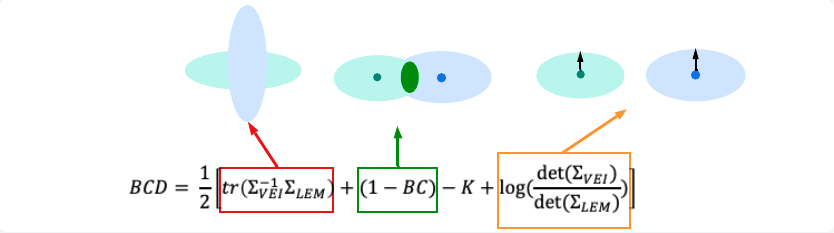}
    \caption{BCD}
    \label{fig:bcd_formula}
\end{subfigure}

    \caption{Comparison of the ECD, KLD, and the proposed BCD VoI object-scoring functions.}
    \label{fig:voi_formula_comparison}
    \vspace{-2mm}
\end{figure}
Fig.~\ref{fig:voi_formula_comparison} compares the ECD, KLD, and proposed BCD VoI functions. The first component, shown in red in Fig.~\ref{fig:bcd_formula}, is the relative covariance term and remains consistent with ECD and KLD. It compares the uncertainty structure of the locally sensed object estimate with the corresponding uncertainty in the VEI, and accounts for the relative sensing accuracy.

The second component, shown in green, is the main modification introduced by BCD. Instead of using the Mahalanobis-distance term from KLD, BCD uses $(1-\mathrm{BC}_{o,k})$, where $\mathrm{BC}_{o,k}$ denotes the Bhattacharyya coefficient between the LEM and VEI object-state distributions~\cite{bhattacharyya1943aspect}. For the Gaussian estimates defined in Eqs.~\eqref{eq:lem_estimate} and~\eqref{eq:vei_estimate}, their densities can be written compactly as $p^{j}_{o,k}(\boldsymbol{x})=\mathcal{N}(\boldsymbol{x};\boldsymbol{\mu}^{j}_{o,k},\boldsymbol{\Sigma}^{j}_{o,k})$, where $j\in\{\mathrm{LEM},\mathrm{VEI}\}$. The Bhattacharyya coefficient is:
\begin{equation}
    \mathrm{BC}_{o,k} = \int_{\mathbb{R}^{K}} \sqrt{p^{\mathrm{LEM}}_{o,k}(\boldsymbol{x}) p^{\mathrm{VEI}}_{o,k}(\boldsymbol{x})}\,\mathrm{d}\boldsymbol{x}, 
    \label{eq:bhattacharyya_coefficient}
\end{equation}
where $K=\dim(\boldsymbol{x})$ is the dimension of the object-state representation used in the VoI calculation. The Bhattacharyya coefficient measures the overlap between two distributions. A value close to one indicates strong overlap and therefore strong agreement, whereas a value close to zero indicates weak overlap and therefore strong disagreement. The term $(1-\mathrm{BC})$ therefore increases when the LEM and VEI disagree, but remains bounded between zero and one. The third component, shown in orange, is the log-determinant ratio between the VEI and LEM covariance matrices, and it remains unchanged relative to ECD and KLD. This term captures the contrast between what the network is believed to know vs what is known locally.

Thus, BCD preserves the covariance-driven behaviour needed for State A and State D, while using the bounded overlap term to separate State B from State C. When the LEM and VEI overlap, $(1-\mathrm{BC})$ remains small; when their overlap is weak, it increases the score without becoming unbounded like KLD. Because the Bhattacharyya coefficient is symmetric, neither the LEM nor the VEI is treated as the reference distribution. For a fixed mean-state separation and comparable covariance shapes, two uncertain estimates overlap more strongly and consequently receive a smaller disagreement contribution. Conversely, two confident but well-separated estimates have little overlap and receive a larger disagreement contribution. BCD therefore recognises disagreement without allowing it to grow unchecked and thus is more conflict-aware than ECD but less aggressive than KLD.

\subsection{Illustrative Object-State Disagreement Case}
\label{subsec:bakld_case_study}
\begin{table*}[t]
\centering
\caption{Means and covariances used in the \textit{Disagreement-aware VoI Scoring} case study (single frame).}
\label{tab:bakld_case_study}
\scriptsize
\setlength{\tabcolsep}{7pt}
\renewcommand{\arraystretch}{1.15}
\begin{tabular}{|c|c|c|}
\hline
\textbf{Object} &
$\boldsymbol{\mu}_{\mathrm{LEM}},~\boldsymbol{\Sigma}_{\mathrm{LEM}}$ &
$\boldsymbol{\mu}_{\mathrm{VEI}},~\boldsymbol{\Sigma}_{\mathrm{VEI}}$ \\
\hline
$O_1$ (VRU) &
$\boldsymbol{\mu}_{\mathrm{LEM}}=[20,\;2.65]^\top,\;
\boldsymbol{\Sigma}_{\mathrm{LEM}}=\mathrm{diag}(3.20,\;1.40)$ &
$\boldsymbol{\mu}_{\mathrm{VEI}}=[20,\;0]^\top,\;
\boldsymbol{\Sigma}_{\mathrm{VEI}}=\mathrm{diag}(2.20,\;2.00)$
\\
\hline
$O_2$ (motorbike) &
$\boldsymbol{\mu}_{\mathrm{LEM}}=[30,\;0]^\top,\;
\boldsymbol{\Sigma}_{\mathrm{LEM}}=\mathrm{diag}(0.25,\;0.25)$ &
$\boldsymbol{\mu}_{\mathrm{VEI}}=[30,\;0]^\top,\;
\boldsymbol{\Sigma}_{\mathrm{VEI}}=\mathrm{diag}(9.00,\;9.00)$
\\
\hline
$O_3$ (vehicle opposite lane) &
$\boldsymbol{\mu}_{\mathrm{LEM}}=[22,\;0]^\top,\;
\boldsymbol{\Sigma}_{\mathrm{LEM}}=\mathrm{diag}(8.00,\;0.90)$ &
$\boldsymbol{\mu}_{\mathrm{VEI}}=[21.70,\;1.35]^\top,\;
\boldsymbol{\Sigma}_{\mathrm{VEI}}=\mathrm{diag}(8.50,\;0.50)$
\\
\hline
\end{tabular}
\vspace{-2mm}
\end{table*}
To illustrate the behaviour of BCD, consider the single-frame scenario in Fig.~\ref{fig:bakld_case_study}. ITS-S A observes three non-connected road users ($O_{1}$-$O_{3}$) and under DCC payload constraints, at most two objects can be included in the outgoing CPM. ITS-S B is the receiver and does not have direct line-of-sight to the vulnerable road user (VRU), denoted by $O_{1}$. The desired behaviour is for ITS-S A to include both an object that is useful for reducing ITS-S B's network-side uncertainty (O2) and an object whose LEM and VEI disagreement may affect a safety-relevant decision (O1). The three objects are defined as follows. Object $O_{1}$ is a VRU with a safety-relevant State B conflict: the LEM and VEI have comparable confidence but place the object at different lateral positions. Object $O_{2}$ is a strong State A object: the LEM and VEI agree on the mean state, but the VEI uncertainty is much larger than the LEM uncertainty, so an additional report is highly useful to the network. Object $O_{3}$ is an intermediate object with lower urgency. Corresponding parameters are shown in Table~\ref{tab:bakld_case_study}. Using ECD, KLD, and BCD to rank the three objects under the same DCC payload constraints with a two object budget gives the following selections:

\begin{figure}
    \centering
    \includegraphics[width=0.9\linewidth]{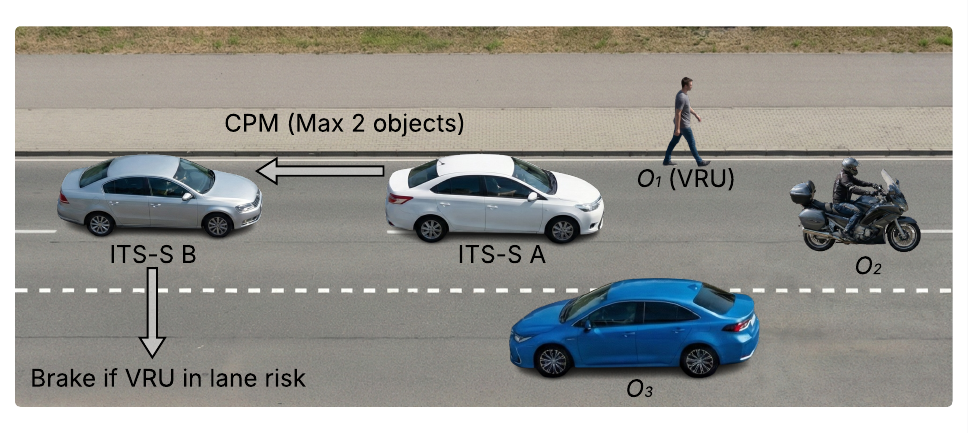}
    \caption{Single-frame snapshot under a DCC payload constraint, where \textit{ITS-S A} can select a maximum of two objects for inclusion in the CPM.}
    \label{fig:bakld_case_study}
    \vspace{-2mm}
\end{figure}

\begin{itemize}
    \item ECD selects $\{O_2(0.72),O_3(0.10)\}$ and omits $O_1(0.06)$ because it only captures the covariance relationship between the LEM and VEI.
    
    \item KLD selects $\{O_2(0.72),O_3(0.66)\}$ and omits $O_1(0.65)$. Although KLD responds to the State B disagreement of $O_1$, its asymmetric covariance weighting assigns a slightly higher score to $O_3$.
    
    \item BCD selects $\{O_2(0.77),O_1(0.32)\}$ and omits $O_3(0.29)$ because its bounded overlap term prioritises the weak LEM--VEI overlap of the State B object while retaining the strong State A object.
\end{itemize}
This snapshot shows the intended behaviour of BCD. ECD is stable but ignores mean-state disagreement. KLD includes disagreement, but its response depends strongly on the VEI covariance and can therefore fail to prioritise the conflict. BCD instead recognises the weak overlap between the LEM and VEI distributions while keeping the disagreement contribution bounded. The single-frame example therefore illustrates one ranking decision under a two-object CPM budget. Temporal simulations next evaluate this selection process over successive CPM generation events and evaluate its cumulative effect.

\section{Simulation Environment}
\label{sec:simulation_environment}
\subsection{Scenario Configuration}
\label{subsec:scenario_configuration}
\begin{figure}[t]
    \centering
    \includegraphics[width=0.8\linewidth]{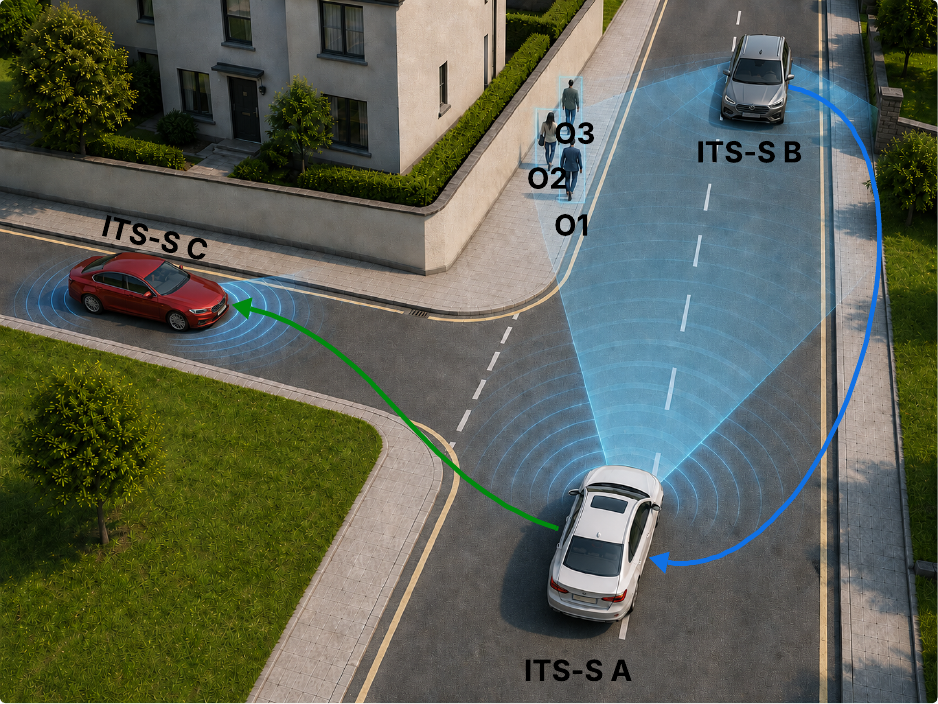}
    \caption{Controlled simulation scenario: The blue arrows represent V2X messages from ITS-S B to ITS-S A, which are used to update the VEI at ITS-S A. The green arrow represents the CPM transmitted from ITS-S A to ITS-S C, carrying the selected object estimates from the LEM of ITS-S A.}
    \label{fig:simulation_scenario}
    \vspace{-2mm}
\end{figure}
Fig.~\ref{fig:simulation_scenario} illustrates a controlled simulation to provide a temporal evaluation of this behaviour. The information flow consists of three stages. First, ITS-S A maintains a LEM using only its onboard sensor observations, while ITS-S B senses the same objects and transmits V2X information to ITS-S A. ITS-S A uses these received reports to update its VEI object trackers. Second, ITS-S A compares the LEM and VEI estimates and ranks the candidate objects using ECD, KLD, or BCD. Third, ITS-S A transmits the selected LEM object estimates to ITS-S C, which cannot directly sense the objects. The receiver-side error at ITS-S C is then used to evaluate the downstream effect of the object-selection decision. The blue arrows in Fig.~\ref{fig:simulation_scenario} denote the V2X messages from ITS-S B that update the VEI at ITS-S A, while the green arrow denotes the CPM transmitted from ITS-S A to ITS-S C. The LEM and VEI inputs, the filtering process, and the CPM object budget are held constant throughout the comparison; only the VoI function used to rank the candidate objects changes. Object $O_1$ is the State B object and represents a potentially safety-relevant LEM--VEI disagreement. Object $O_2$ is configured as a strong State A object, where the LEM is more certain than the VEI. Object $O_3$ is used as a lower-priority reference object, corresponding to a weak State A or State C-like case. This evaluation determines which VoI function recognises disagreement while continuing to allocate transmission opportunities to other useful objects.

\subsection{Simulation Modelling}
\label{subsec:simulation_setup}
The simulation runs for $100$ time steps of $\Delta t=0.1$~s, corresponding to a duration of $10$~s and an update frequency of $10$~Hz. Both the LEM and VEI are maintained using standard linear Kalman filters with the four-dimensional constant-velocity state ${[x,\,y,\,v_x,\,v_y]}^{\mathsf{T}}$. The Kalman filter configuration used to generate the LEM and VEI estimates and determine how the three object states are created are summarised in Table~\ref{tab:state_noise_model}. Their roles are summarised as follows:

\begin{table}[t]
\centering
\caption{Kalman filter state and noise models.}
\label{tab:state_noise_model}
\renewcommand{\arraystretch}{1.15}
\resizebox{\linewidth}{!}{
\begin{tabular}{|p{0.40\linewidth}|>{\raggedright\arraybackslash}p{0.44\linewidth}|}
\hline
\textbf{Parameter} & \textbf{Value} \\
\hline
Kalman filter state
& ${[x,\,y,\,v_x,\,v_y]}^{\mathsf{T}}$ \\
\hline
Process noise $\mathbf{Q}$
& $\mathrm{diag}[0.02,\,0.02,\,0.05,\,0.05]$ \\
\hline
ITS-S A sensor noise
& $\mathrm{diag}[0.30^2,\,0.30^2,\,0.20^2,\,0.20^2]$ \\
\hline
ITS-S B sensor noise
& $\mathrm{diag}[0.35^2,\,0.35^2,\,0.20^2,\,0.20^2]$ \\
\hline
ITS-S B bias on $O_1$
& $[0.0,\,2.5,\,0.0,\,0.0]^{\mathsf{T}}$ \\
\hline
Initial velocity covariance
& LEM and VEI: $\mathrm{diag}[0.20^2,\,0.20^2]$ \\
\hline
$O_1$ initial position covariance
& LEM: $\mathrm{diag}[0.50^2,\,0.50^2]$;
  VEI: $\mathrm{diag}[0.53^2,\,0.53^2]$ \\
\hline
$O_2$ initial position covariance
& LEM: $\mathrm{diag}[0.35^2,\,0.35^2]$;
  VEI: $\mathrm{diag}[2.00^2,\,2.00^2]$ \\
\hline
$O_3$ initial position covariance
& LEM: $\mathrm{diag}[0.80^2,\,0.80^2]$;
  VEI: $\mathrm{diag}[1.20^2,\,1.20^2]$ \\
\hline
\end{tabular}
}
\end{table}

The sensor-noise settings control the LEM and VEI uncertainty, while the bias applied to the V2X-derived estimate of $O_1$ creates the State~B disagreement. The initial covariances configure $O_1$, $O_2$, and $O_3$ as the disagreement, strong State~A, and lower-priority reference objects, respectively. At each CPM generation event, ITS-S A ranks the candidate objects using ECD, KLD, or BCD and transmits only the highest-ranked objects permitted by the DCC payload constraint. Performance is evaluated using three metrics: (i) the receiver-side estimation error for each object; (ii) the aggregate error across $O_1$, $O_2$, and $O_3$; and (iii) the proportion of CPM generation events in which each object is selected. Together, these metrics show whether reducing the State~B object error comes at the cost of neglecting other objects in the receiver-side environment model.

\section{Results and Evaluation}
\label{sec:results}
This section compares ECD, KLD, and BCD under the controlled State B conditions described in Section~\ref{subsec:simulation_setup}. Results first examine object inclusion under DCC payload constraints in Section~\ref{subsec:budget_sensitivity}, and then test robustness when the local LEM estimate is biased in Section~\ref{subsec:reversed_bias}.

\subsection{Budget Sensitivity under DCC payload Constraints}
\label{subsec:budget_sensitivity}
\begin{figure}[t]
    \centering
    \includegraphics[width=\linewidth]{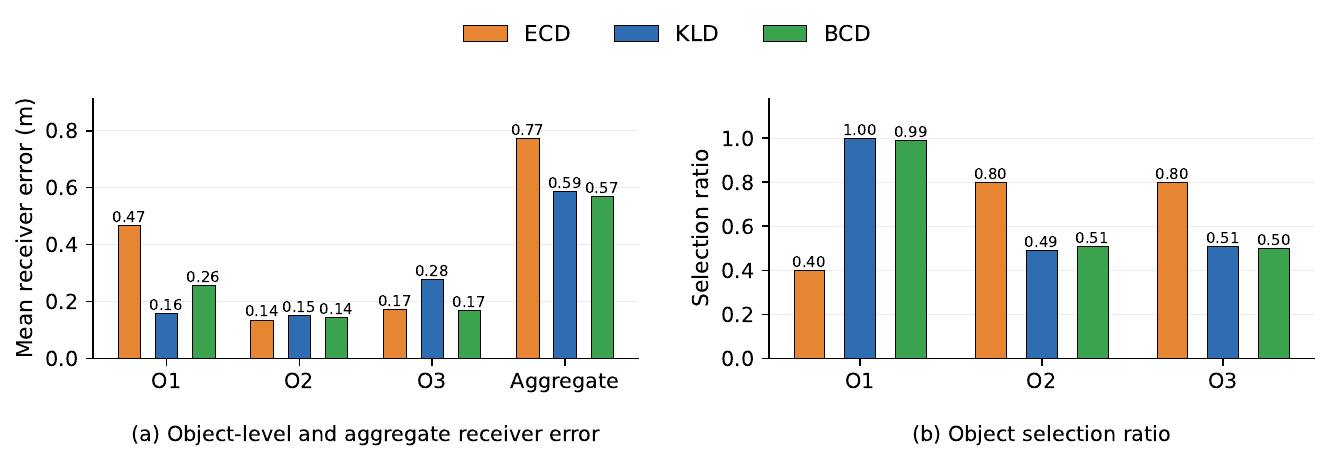}
    \caption{Under a \textbf{2 object} DCC payload constraint, this shows \textbf{receiver-side} (ITS-S C) (a) object-level and aggregate error, and (b) object selection ratio.}
    \label{fig:topk2_combined}
    \vspace{-2mm}
\end{figure}

\begin{figure}[t]
    \centering
    \includegraphics[width=\linewidth]{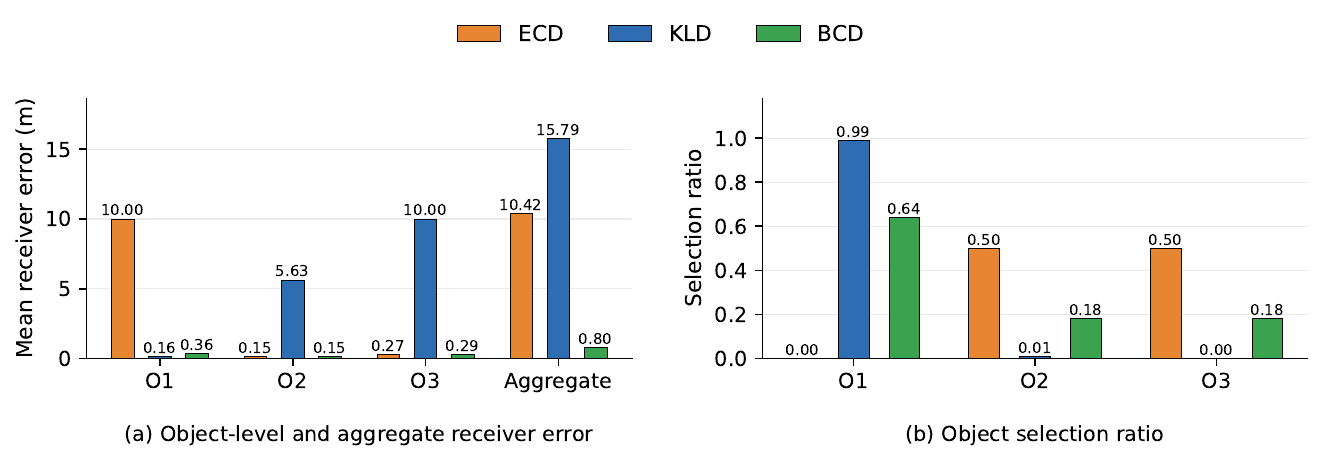}
    \caption{Under a \textbf{1 object} DCC payload constraint, this shows \textbf{receiver-side} (ITS-S C) (a) object-level and aggregate error, and (b) object selection ratio.}
    \label{fig:topk1_combined}
    \vspace{-2mm}
\end{figure}

Fig.~\ref{fig:topk2_combined} shows the receiver-side error and object selection ratio under a constraint where ITS-S A can include at most two objects in each CPM. In contrast, Fig.~\ref{fig:topk1_combined} considers a more restrictive DCC payload constraint with a maximum of 1 object allowed for inclusion within the CPM. These two settings allow us to analyse how each VoI function behaves as the DCC payload constraint becomes more restrictive. The purpose of this evaluation is not to show that BCD produces the lowest receiver error for every object individually. Prioritising one object consumes part of the limited CPM object budget and can therefore increase the error of another object. The principal comparison is the aggregate receiver error, considered together with the individual object errors and selection ratios. Accordingly, BCD can provide the best overall balance even when ECD or KLD gives a lower error for a particular object.
It can be observed in Fig.~\ref{fig:topk2_combined} that ECD gives low mean receiver errors for $O_{2}$ and $O_{3}$, but produces a higher error for $O_{1}$. This is expected because ECD is driven by the LEM and VEI covariance relationship only and does not directly account for the mean-state disagreement of the State B object. KLD reduces the error of $O_{1}$, showing that its Mahalanobis-distance term reacts to the disagreement between the LEM and VEI. However, this comes with a higher error for $O_{3}$, which evidences that a stronger focus on the disagreement object can result in exclusion of other objects, thereby reducing the quality of the wider receiver-side environment model. BCD provides the lowest aggregate receiver error. It reduces the error of $O_{1}$ compared with ECD, while avoiding the larger degradation of $O_{3}$ observed with KLD. The selection ratios in Fig.~\ref{fig:topk2_combined}(b) help explain this behaviour. ECD selects $O_{1}$ less frequently than the other two functions, because it does not directly recognise the State B disagreement. KLD and BCD both select $O_{1}$ in almost every CPM. However, after $O_{1}$ is selected, the remaining object position is distributed differently between $O_{2}$ and $O_{3}$. This shows that the improvement in BCD is not solely due to selecting $O_{1}$ more often. It also comes from a more balanced ranking behaviour, which preserves useful updates for the other objects.

Fig.~\ref{fig:topk1_combined} shows the impact of the stricter DCC payload constraint of one object. Here, the difference between the three functions becomes more pronounced. ECD never selects $O_{1}$ and therefore produces a very high receiver-side error for the State B object. This confirms that a covariance-only VoI function can miss a potentially safety-relevant disagreement when the LEM and VEI have comparable uncertainty. KLD shows the opposite behaviour. It selects $O_{1}$ in almost every CPM, which strongly reduces the $O_{1}$ error, but this causes $O_{2}$ and $O_{3}$ to receive almost no updates. As a result, KLD gives the highest aggregate receiver error under the strict budget. BCD provides the best compromise under such restrictive DCC payload constraints. It selects $O_{1}$ most often, which keeps the State B error much lower than ECD. However, it still reserves part of the CPM budget for $O_{2}$ and $O_{3}$, avoiding the object starvation observed with KLD. This leads to the lowest aggregate receiver error among the evaluated functions. 

\subsection{Reversed-Bias Robustness Evaluation}
\label{subsec:reversed_bias}
\begin{figure}[t]
    \centering
    \includegraphics[width=\linewidth]{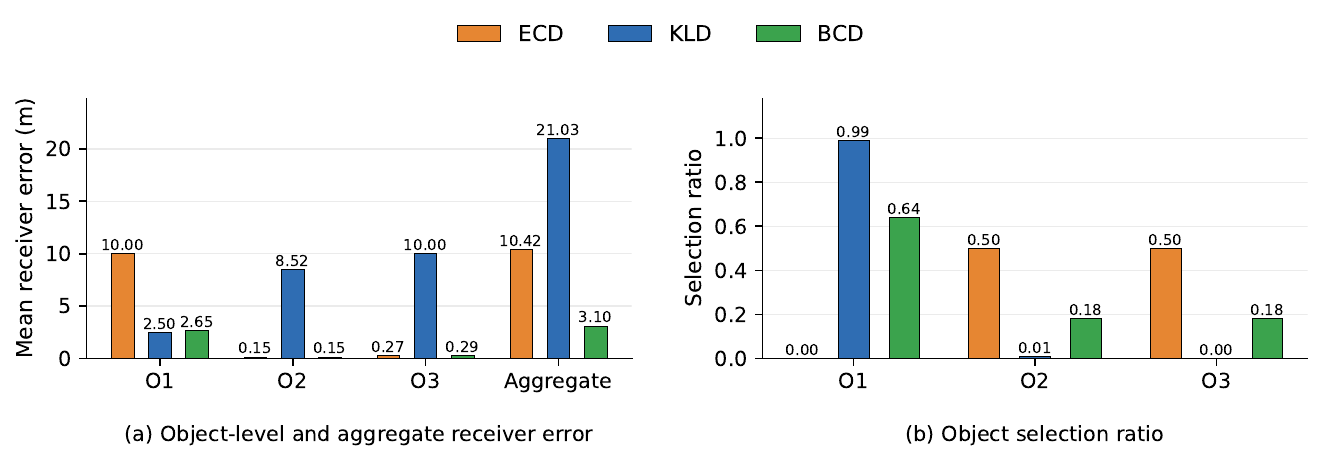}
    \caption{Receiver-side error and object selection ratio under the reversed-bias test with a DCC payload constraints with one object budget. In this case, the LEM estimate of $O_{1}$ is biased, while the VEI is closer to the true object state.}
    \label{fig:reversed_bias_combined}
    \vspace{-2.0mm}
\end{figure}
As discussed in Section~\ref{subsec:state_b_causes}, State B indicates LEM and VEI inconsistency, but not which estimate is correct. Thus we examine whether a disagreement-aware VoI function propagates erroneous local information when the LEM estimate is biased, which could mislead downstream decisions. A strict one-object DCC payload constraint is used, with the LEM estimate of $O_1$ treated as biased and the VEI closer to the true object state. This reverses the previous assumption and evaluates whether each VoI function continues to prioritise $O_1$ and how this affects the receiver-side model at ITS-S C.

Fig.~\ref{fig:reversed_bias_combined} shows that ECD does not select $O_{1}$ because it does not explicitly score the mean-state disagreement. This avoids repeatedly transmitting the biased LEM estimate of $O_{1}$. However, it also means that ITS-S C receives few useful updates about $O_{1}$, resulting in a high $O_{1}$ error. KLD shows the opposite behaviour. It selects $O_{1}$ in almost every CPM because the Mahalanobis-distance term strongly reacts to the LEM and VEI disagreement. This reduces the $O_{1}$ error compared with ECD, but the transmitted estimate remains biased because the LEM itself is incorrect. More importantly, KLD leaves $O_{2}$ and $O_{3}$ with almost no update opportunities. As a result, it produces the highest aggregate receiver error. This shows that an aggressive disagreement-aware function can be harmful when the local estimate is not the correct one. BCD provides the most robust overall behaviour. It still recognises the disagreement and selects $O_{1}$ frequently, but it does not allow $O_{1}$ to monopolise the available object position. This preserves update opportunities for $O_{2}$ and $O_{3}$ and leads to the lowest aggregate receiver error. Overall, the reversed-bias test highlights a key safety consideration. Disagreement is valuable because it reveals an inconsistency in the collective perception state. However, if the VoI function responds too aggressively, it may repeatedly propagate erroneous information. 

\section{Conclusion and Future Work}
\label{sec:conclusion}
This paper proposes BCD, a bounded disagreement-aware CPS VoI function that specifically addresses edge cases where the local and network estimates have comparable uncertainty but disagree in the estimated object state. The results show that BCD provides a balanced treatment of disagreement under constrained CPM budgets, producing the lowest aggregate receiver error in the evaluated scenarios. Under stricter DCC payload constraints, this bounded response becomes particularly important, as KLD can allow the State B object to dominate and degrade the wider receiver-side environment model. It also showed that when the local estimate is erroneous, an overly aggressive disagreement-aware function can repeatedly propagate erroneous information to the receiver. BCD mitigates this effect by recognising the disagreement while limiting its dominance. 

\section*{Acknowledgment}
This publication has emanated from research conducted with the financial support of Research Ireland under 18/CRT/6222 \& 13/RC/2077 P2. The author has applied a CC BY public copyright licence to any Author Accepted Manuscript version arising from this submission.

\bibliography{references}
\bibliographystyle{IEEEtran}
\end{document}